\documentstyle[12pt,a4wide]{article}
\newcommand\GeV{{\,\sl GeV}}
\newcommand\Frac[2]{\mbox{$\textstyle{#1\over#2}$}}
\newcommand\acknowledgements{\vspace{0.5truecm}\noindent{\bf 
  Acknowledgements:}\ }

\begin{document}

\thispagestyle{empty}
\begin{flushright}
MZ-TH/97-09\\
hep-ph/9703208\\
March 1997\\
\end{flushright}
\vspace{0.5cm}
\begin{center}
{\Large\bf New high order relations between}\\[.2truecm]
{\Large\bf physical observables in perturbative QCD}\\[1truecm]
{\large S.~Groote,$^1$ J.G.~K\"orner,$^1$ A.A.~Pivovarov,$^{1,2}$
  and K.~Schilcher$^1$}\\[.7cm]
$^1$ Institut f\"ur Physik, Johannes-Gutenberg-Universit\"at,\\[.2truecm]
  Staudinger Weg 7, D-55099 Mainz, Germany\\[.5truecm]
$^2$ Institute for Nuclear Research of the\\[.2truecm]
  Russian Academy of Sciences, Moscow 117312
\vspace{1truecm}
\end{center}

\begin{abstract}
We exploit the fact that within massless perturbative QCD the same Green's 
function determines the hadronic contribution to the $\tau$ decay width and 
the moments of the $e^+e^-$ cross section. This allows to obtain relations
between physical observables in the two processes up to an unprecedented 
high order of perturbative QCD. An precision measurement of the $\tau$ 
decay width allows one then to predict the first few moments of the 
spectral density in $e^+e^-$ annihilations integrated up to 
$s\sim m_\tau^2$ with high accuracy. The proposed tests are in the reach 
of present experimental capabilities.
\end{abstract}

\newpage

Due to the availability of ever-increasing orders of perturbative QCD 
results on the one hand and continuing improvements on the quality of 
experimental data on the other hand, high order precision test of the QCD 
part of the Standard Model are now becoming feasible~\cite{altrev}. In 
contrast to the electroweak sector where the expansion parameter for the 
perturbative calculations is sufficiently small and one or two corrections 
provide a theoretical accuracy sufficient for a comparison with present 
experiments, the situation with the strong interactions is different, 
maybe even in principle. The relevant expansion parameter is large, such 
that a desired accuracy of order $1\%$ requires four- to five-loop accuracy 
in the perturbative expansion. This is close to the border of today's 
computational possibilities. In addition, with such a large number of 
terms, one may already encounter the asymptotic nature of the perturbation 
series in which case no further increase of precision is possible in 
straightforward perturbation theory.

The above question of numerical convergence is influenced to a large extent 
by the renormalization scheme dependence of the truncated perturbation 
series~\cite{Grunberg,Stevenson,Brodsky}. Therefore it is desirable to 
obtain predictions for observables which are renormalization scheme 
independent~\cite{Nordita}.

In this letter we propose a new test of {\em perturbative QCD\/} that 
is free from renormalization scheme ambiguities and is of higher precision 
in terms of the number of corrections than any previous test. This test 
directly expresses one observable through another observable using massless 
perturbative QCD as the underlying theory. A disagreement with experimental 
data would imply a failure of the applicability of perturbative QCD.

The advocated comparison is made for the moments of the spectral density 
in $e^+e^-$ annihilation and the hadronic contributions to 
$\Gamma(\tau\rightarrow\nu_\tau+{\rm hadrons})$. Within massless perturbative 
QCD these quantities have been computed up to four-loop order. We define 
a reduced semileptonic $\tau$ decay ratio $r_\tau$ through
\begin{equation}\label{eqn1}
R_\tau=\frac{\Gamma(\tau\rightarrow\nu_\tau+{\rm hadrons})}%
  {\Gamma(\tau\rightarrow\nu_\tau+\mu+\bar\nu_\mu)}=R_\tau^0(1+r_\tau)
\end{equation}
where $R_\tau^0$ is the lowest order partonic value of $R_\tau$. To lowest 
order in the electroweak interactions and assuming 
$|V_{ud}|^2+|V_{us}|^2\cong 1$ it takes the value $R_\tau^0=N_c=3$.
Corrections due to higher order electroweak interactions and power 
suppressed terms are small~\cite{BraatenNarisonPich,Pich} and can be easily 
accounted for by rescaling the partonic ratio $R_\tau^0$. Thus the reduced 
decay width $r_\tau$ is determined by massless perturbative QCD, for which 
the axial and vector contributions are identical. The expansion for 
$r_\tau$ starts directly with $a(\mu^2)$, where $a=\alpha_s/\pi$. The 
dependence of $a(\mu^2)$ on the renormalization scale~$\mu$ is determined 
by the renormalization group equation
\begin{equation}
\mu^2\frac{da}{d\mu^2}=\beta(a)=-a^2(\beta_0+\beta_1a+\beta_2a^2
  +\beta_3a^3+\ldots\ ).
\end{equation}
In minimal subtraction schemes one has
\begin{eqnarray}
\beta_0&=&\frac94=2.25,\nonumber\\
\beta_1&=&\frac{64}{16}=4,\nonumber\\
\beta_2&=&\frac{3863}{384}\approx 10.060,\nonumber\\
\beta_3&=&\frac{140599}{4608}+\frac{445}{32}\zeta(3)\approx 47.228
\end{eqnarray}
for $N_c=n_f=3$. Here the recently computed four-loop coefficient 
$\beta_3$~\cite{Ritbergen} is also included. In $e^+e^-$ annihilation the 
cross section is determined by the imaginary part of the vacuum 
polarization,
\begin{equation}
R_{e^+e^-}(s)=12\pi{\sl Im\,}\Pi(s)=N_c\sum Q_i^2(1+r(s))=2(1+r(s)).
\end{equation}
In perturbative QCD one has
\begin{eqnarray}
r(s)&=&a(\mu^2)+(k_1+\beta_0L)a^2(\mu^2)+\Big(k_2-\Frac13\pi^2\beta_0^2
  +(2\beta_0k_1+\beta_1)L+\beta_0^2L^2\Big)a^3(\mu^2)\nonumber\\&&
  +\Big(k_3-\pi^2\beta_0^2k_1-\Frac56\pi^2\beta_0\beta_1
  +(3\beta_0k_2+2\beta_1k_1+\beta_2-\pi^2\beta_0^3)L\nonumber\\&&
  +(3\beta_0k_1+\Frac52\beta_1)L^2+\beta_0^3L^3\Big)a^4(\mu^2)+\ldots
\end{eqnarray}
with $L=\ln(\mu^2/s)$ and
\begin{eqnarray}
k_1&=&\frac{299}{24}-9\zeta(3)\approx 1.63982,\nonumber\\
k_2&=&\frac{58057}{288}-\frac{779}4\zeta(3)+\frac{75}2\zeta(5)\approx 6.37101
\end{eqnarray}
in the modified minimal subtraction ($\overline{\rm MS}$) 
scheme~\cite{Gorishny}, while $k_3$ is still unknown.

We can define moments of the reduced part $r(s)$ of the spectral density,
\begin{equation}
r_n(s_0)=(n+1)\int_0^{s_0}\frac{ds}{s_0}\left(\frac{s}{s_0}\right)^nr(s)
\end{equation}
which, for the sake of convenience, are normalized to unity for $r(s)=1$. 
In terms of these moments the reduced decay width ratio $r_\tau$ is given by
\begin{equation}\label{eqn2}
r_\tau=2r_0(m_\tau^2)-2r_2(m_\tau^2)+r_3(m_\tau^2).
\end{equation}
Eq.~(\ref{eqn2}) can be inverted within perturbation theory. One can then 
express the perturbative representation of one observable, i.e.\ any given 
$e^+e^-$ moment $r_n(m_\tau^2)$, in powers of $r_\tau$ using the 
perturbative expansion of the $\tau$ decay observable. The strong coupling 
constant $\alpha_s$ in any given scheme ($\overline{\rm MS}$ for the 
present letter) serves only as an intermediate agent to obtain relations 
between physical observables. The reexpression of one perturbative 
observable through another is a perfectly legitimate procedure in 
perturbation theory and is standard practice in electroweak theory. The 
result is independent of the choice of the renormalization scheme. Solving 
for the moments $r_n(m_\tau^2)$ in terms of the reduced $\tau$ decay width 
$r_\tau$ one finds
\begin{equation}
r_n(m_\tau^2)=f_{0n}r_\tau+f_{1n}r_\tau^2+f_{2n}r_\tau^3
  +f_{3n}r_\tau^4+f_{4n}r_\tau^5+O(r_\tau^6),\label{eqn3}
\end{equation}
where
\begin{eqnarray}\label{eqn4}
f_{0n}&=&\tilde I(0,n),\nonumber\\[7pt]
f_{1n}&=&\beta_0\tilde I(1,n),\nonumber\\[3pt]
f_{2n}&=&\beta_0^2\Big(\tilde I(2,n)+\rho_1\tilde I(1,n)\Big),\nonumber\\
f_{3n}&=&\beta_0^3\Big(\tilde I(3,n)+(I_\tau(2)-I_\tau(1)^2-\Frac13\pi^2)
  \tilde I(1,n)+\Frac52\rho_1\tilde I(2,n)+\rho_2\tilde I(1,n)\Big),
  \nonumber\\
f_{4n}&=&\beta_0^4\Big(\tilde I(4,n)-3(I_\tau(2)-I_\tau(1)^2-\Frac13\pi^2)
  \tilde I(2,n)\nonumber\\&&\qquad+2(I_\tau(3)-3I_\tau(1)I_\tau(2)
  +2I_\tau(1)^3)\tilde I(1,n)\\&&
  +\rho_1(\Frac{13}3\tilde I(3,n)+5(I_\tau(2)-I_\tau(1)^2-\Frac13\pi^2)
  \tilde I(1,n))+3\rho_2\tilde I(2,n)+\rho_3\tilde I(1,n)\Big)\nonumber
\end{eqnarray}
with
\begin{eqnarray}
I(m,n)&=&\frac{m!}{(n+1)^m},\nonumber\\
I_\tau(m)&=&2I(m,0)-2I(m,2)+I(m,3),\nonumber\\
I(m,n)&=&I(m)+\sum_{p=0}^m{m\choose p}I_\tau(p)\tilde I(m-p,n).\label{eqn5}
\end{eqnarray}
The coefficients $\tilde I(m,n)$ used in Eqs.~(\ref{eqn4}) can be determined 
iteratively by solving Eq.~(\ref{eqn5}). In general, the coefficient 
functions $f_{in}$ depend on $s_0/m_\tau^2$, but here we limit our analysis 
to the value $s_0=m_\tau^2$. The $\rho_i$ are scheme independent quantities 
given by
\begin{eqnarray}
\rho_1&=&\frac{\beta_1}{\beta_0^2}\approx 0.79012,\nonumber\\
\rho_2&=&\frac1{\beta_0^3}\big[\beta_2-\beta_1k_1+\beta_0(k_2-k_1^2)\big]
  \approx 1.03463,\nonumber\\
\rho_3&=&\frac1{\beta_0^4}\big[\beta_3-2\beta_2k_1+\beta_1k_1^2
  +2\beta_0(k_3-3k_1k_2+2k_1^3)\big]\approx0.17558k_3-2.9795.
\end{eqnarray}
The coefficient $f_{4n}$ contains the unknown five-loop coefficient $k_3$. 
Later on we present an estimate of the coefficient $k_3$ in order to 
be able to assess the importance of the $r_\tau^5$ contributions to the 
perturbation series.

We now turn to the numerical analysis. From the experimental value 
$R_\tau^{\rm exp}=3.649\pm0.014$ (as a recent review see e.g.~\cite{Pich}) 
for the semileptonic $\tau$ decay ratio one gets 
$r_\tau^{\rm exp}=0.216\pm 0.005$. This value is obtained from 
Eq.~(\ref{eqn1}) using $R_\tau^0=3$. Nonperturbative and electroweak 
corrections lie within the error bars. In order to investigate the 
convergence properties of the series we calculate the first few moments 
by inserting the numerical values of all known coefficients in 
Eq.~(\ref{eqn3}). One has
\begin{eqnarray}
r_{-1/2}&=&0.216(1+0.203+0.786+2.637
  +(10.601+0.0040k_3))+O(r_\tau^6),\label{eqn6}\\
r_0&=&0.216(1-0.284-0.069+0.110+(0.211-0.0057k_3))+O(r_\tau^6),\\
r_1&=&0.216(1-0.527-0.143+0.177+(0.361-0.0106k_3))+O(r_\tau^6),\\
r_2&=&0.216(1-0.608-0.115+0.269+(0.433-0.0122k_3))+O(r_\tau^6),\\
r_3&=&0.216(1-0.648-0.091+0.317+(0.444-0.0131k_3))+O(r_\tau^6).
\end{eqnarray}
We have started with the moment $r_{-1/2}$ because it represents the cross 
section integrated over the energy $\sqrt{s}$ which would be the natural 
choice of moment when considering the usual representation of the 
experimental data. It is quite apparent from Eq.~(\ref{eqn6}) that the 
series for $n=-1/2$ does not converge at all. A possible explanation is 
that the low energy region of the spectral function is emphasized too much 
for perturbation theory to work properly.

To discuss the convergence properties of the perturbation series for the 
other moments ($n=0,1,2,3$) we distinguish two scenarios depending on how 
the unknown $r_\tau^5$ term is accounted for. These are
\renewcommand{\labelenumi}{(\alph{enumi})}
\begin{enumerate}
\item Nothing is known about the $r_\tau^5$ term (i.e.\ about $k_3$). 
Then we truncate the series at the smallest term and consider the next term 
as an estimate of the error. We call this the {\em conservative estimate}.
\item We adopt a value $k_3=k_2^2/k_1\approx 25$ as estimated by using 
Pad\'e techniques~\cite{LeDiberderPich,EllisKarlinerSamuel}. There are 
other methods to estimate $k_3$ which give results close to the Pad\'e 
estimate~\cite{KataevStarshenko}. With this choice of $k_3$ the series 
appears to consist of two alternating series, one of them with large terms 
and the other with small terms. With such a structure our best prediction 
lies between the third order and the fourth order partial sum, so the error 
is half of the difference of the two. Note that the correction to the 
fourth order partial sum is always small. We call this the {\em educated 
estimate}.
\end{enumerate}
We collect our numerical results for the full moments $R_n=2(1+r_n)$ 
and their errors in Table~\ref{tab1}.
\begin{table}[t]\begin{center}
\begin{tabular}{|r|c|c|c|}\hline
  &(a)&(b)&(c)\\\hline
  $R_0$&$2.28\pm 0.05$&$2.30\pm 0.02$&$2.15$\\
  $R_1$&$2.14\pm 0.08$&$2.18\pm 0.04$&$2.06$\\
  $R_2$&$2.12\pm 0.12$&$2.18\pm 0.06$&$2.00$\\
  $R_3$&$2.11\pm 0.14$&$2.18\pm 0.07$&$1.99$\\\hline
\end{tabular}
\caption{Numerical values of the $e^+e^-$ moments for the conservative 
estimate (a), for the educated estimate (b), and for the experimental 
$e^+e^-$ moments (c).\label{tab1}}
\end{center}\end{table}
Here we also list results for the experimental moments using the 
experimental $e^+e^-$ data from a recent compilation~\cite{Eidelman}. On 
average the error bars of the conservative estimate (a) are a factor two 
bigger than those of the educated estimate (b). For $R_0$ the error is 
very small (even for the conservative estimate) because the perturbation 
series is reliable and converges well numerically. The data seem to be 
systematically lower by about 7\%. The resulting statistical error for the 
moments of the data is negligible, the systematic error could be as large 
as 10\%~\cite{Eidelman,Dolinsky}. For the higher moments the theoretical 
predictions are less accurate. The experimental values lie within the error 
bars of the conservative estimate (a). If one takes the educated estimate 
(b), the experimental values are consistently lower than the theoretical 
prediction. If the data points were increased by 7\%, the moments would be 
consistent with both theoretical predictions.

We have also checked on the non-perturbative contributions to the moments 
$R_0$, $R_1$ and $R_2$. The non-perturbative contribution to $R_0$ vanishes, 
since there is no gauge invariant operator of dimension two. The 
non-perturbative contributions to $R_1$ and $R_2$ are well within the 
truncation error of the perturbation series.

As a last point we consider the relation in Eq.~(\ref{eqn2}) which holds to 
arbitrary order in perturbative QCD. If we use experimental data on both 
sides of Eq.~(\ref{eqn2}), we find the value $3.449$ for the right hand 
side, so the compiled data of the $e^+e^-$ annihilation are lower by 6\% 
in comparison with $r_\tau$. On the other hand, if we use the predicted 
moments of Table~\ref{tab1} to evaluate the right hand side of 
Eq.~(\ref{eqn2}), we obtain $3.648$ in both cases (a) and (b). This is very 
close to the input value $r_\tau=3.649\pm 0.005$ and within our truncation 
errors which means that our procedure is self-consistent. This observation 
confirms the basic result of perturbative QCD that moments of vector and 
axial vector spectral functions are equal to one another. If against all 
expectations future precise $e^+e^-$ data does not satisfy Eq.~(\ref{eqn2}) 
then this would imply that the moments of the vector and axial vector 
spectral functions differ and that perturbative QCD is in trouble at the 
scale of $m_\tau^2$.

In summary we have established perturbative relations between different 
sets of observables with one more term in the perturbation series than ever 
considered before. This seemingly surprising result can be traced to the 
fact that the observables are generated from the same Green's function. The 
coefficients $f_{in}$ in our perturbative expansion are scheme independent 
quantities. The stated relations between the observables are quite useful 
and can be tested since the observables can be measured directly, 
independently and very precisely.

Our analysis could be extended in two directions as concerns possible 
improvements on the convergence properties of the series expansion. One can 
either vary the endpoint $s_0$ of the moment integration in order to 
optimize the series expansion in Eq.~(\ref{eqn3}) or one can attempt to 
find an optimal linear combination of moments that has improved convergence 
properties without a concomitant loss of precision through cancellation 
effects.

In view of the fact that the $e^+e^-$ cross section will be remeasured with 
considerably improved accuracy at ${\rm DA\Phi NE}$~\cite{Daphne} we suggest 
the following treatment of the data. For our analysis only integrals of the 
data are needed. Excluding the resonance region (say $E<1\GeV$) the 
resulting spectral density is sufficiently smooth because the integral is a 
linear combination of data points. It is therefore sufficient to have only 
a few points but measured with very high precision.

\acknowledgements
This work is partially supported by the BMBF, FRG, under contract 
No.~06MZ566. A.A.P. greatfully acknow\-ledges partial financial support 
by the Russian Fund for Basic Research under contract No.~96-01-01860. The 
work of S.G. is supported by the DFG, FRG.

\end{document}